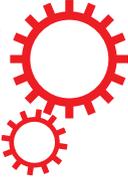

# Magnetism in Na-filled Fe-based skutterudites

Guangzong Xing[1], Xiaofeng Fan[1], Weitao Zheng[1], Yanming Ma[2], Hongliang Shi[3] & David J. Singh[3]

The interplay of superconductivity and magnetism is a subject of ongoing interest, stimulated most recently by the discovery of Fe-based superconductivity and the recognition that spin-fluctuations near a magnetic quantum critical point may provide an explanation for the superconductivity and the order parameter. Here we investigate magnetism in the Na filled Fe-based skutterudites using first principles calculations. $NaFe_4Sb_{12}$ is a known ferromagnet near a quantum critical point. We find a ferromagnetic metallic state for this compound driven by a Stoner type instability, consistent with prior work. In accord with prior work, the magnetization is overestimated, as expected for a material near an itinerant ferromagnetic quantum critical point. $NaFe_4P_{12}$ also shows a ferromagnetic instability at the density functional level, but this instability is much weaker than that of $NaFe_4Sb_{12}$, possibly placing it on the paramagnetic side of the quantum critical point. $NaFe_4As_{12}$ shows intermediate behavior. We also present results for skutterudite $FeSb_3$, which is a metastable phase that has been reported in thin film form.

An important feature of the Fe-based superconductors and the related antiferromagnetic phases is that the energy scale for the spin fluctuations is quite high, for example with spin wave dispersions extending well over 100 meV, hardly indicative of materials in which the magnetism is readily suppressed[1–11]. Nonetheless, this is the case and in particular the magnetic orderings can generally be readily suppressed in favor of superconductivity by doping, isovalent alloying and pressure.

One scenario is that the ordering is suppressed by quantum fluctuations. In this case, at the mean field level the compounds would be strongly magnetic over the entire range of the phase diagrams, similar to what has been discussed near magnetic quantum critical points[12,13]. For example, standard density functional theory (DFT) calculations (done with the PBE generalized gradient approximation and the experimental crystal structure) predict ordered magnetism with moments of $\sim 2\,\mu_B$ over the entire composition range of $LaFeAs(O,F)$[14], while in fact the ordered moment is $\sim 0.5\,\mu_B$ for $LaFeAsO$ and is rapidly suppressed by fluorine doping[8,9]. This type of overestimate of magnetic tendencies within density functional calculations is unusual. It typically arises when spin fluctuations associated with a nearby quantum critical point are strong enough to renormalize the mean-field like magnetic state predicted by standard approximate density functional calculations[15,16]. Thus these results suggest the presence of strong renormalization by quantum fluctuations in the Fe-based superconductors. These would presumably include longitudinal fluctuations[17], recently observed in $BaFe_2As_2$[18]. This is supported by experimental measurements for example the observation by xray-absorption spectroscopy of a large exchange splitting of the Fe 3s state in superconducting $CeFeAs(O,F)$ as if the compound were magnetically ordered, even though it is not magnetic at the composition studied[19]. This is qualitatively similar to but stronger than what is seen in $NbFe_2$, which is known to be close to a quantum critical point[20–22].

Turning to the chemistry of the Fe-based superconductors, all the phases have layers of Fe in a formal divalent or close to divalent state and tetrahedral coordination by the pnictogens, P or As, or Se or alloys with Te and/or S. There has been theoretical work suggesting stronger magnetism and possibly superconductivity in hypothetical Fe antimonide compounds analogous to the superconducting arsenides[23], but to date these compounds have not been synthesized. While there are many other families of Fe compounds that show magnetism, and in some cases the magnetism can be suppressed by alloying or pressure, in general these other compounds do not display the interesting magnetic properties of the Fe-based superconductors. Nonetheless, it may be that the strong renormalizations present in the Fe-based superconductors may not be unique to those compounds and it is of considerable interest to find other iron compounds displaying such behavior.

[1]College of Materials Science and Engineering, Jilin University, 130012, Changchun, China. [2]State Key Lab of Superhard Materials, Jilin University, 130012, Changchun, China. [3]Materials Science and Technology Division, Oak Ridge National Laboratory, Oak Ridge, TN 37831-6056, USA. Correspondence and requests for materials should be addressed to D.J.S. (email: singhdj@ornl.gov)

Recently, calculations suggest that such a suppression of magnetic order by quantum fluctuations with qualitatively similar size to the Fe-based superconductors can occur in the Ge compound, $YFe_2Ge_2$[24,25]. This is a compound that shows other indications of nearness to a magnetic quantum critical point, and also possibly low temperature superconductivity[26,27]. The electronic structure of this compound is considerably more three dimensional than any of the Fe-based superconductors. It is therefore of interest to search for other compounds that might show related physics.

Filled skutterudites are a large family of transition metal pnictide compounds that have been extensively studied in the context of thermoelectrics, but also show other interesting behaviors including Fe-based magnetism in $NaFe_4Sb_{12}$[28–30], and both conventional (e.g. $LaRu_4P_{12}$)[31] and unconventional superconductivity (e.g. $PrOs_4Sb_{12}$)[32,33].

Depending on the filling atom, the filled Fe-based skutterudites form with all of P, As and importantly Sb (e.g. $LaFe_4P_{12}$, $LaFe_4As_{12}$, $LaFe_4Sb_{12}$). This is unlike the Fe-based superconducting structures, for which antimonides do not exist. This provides an opportunity for studying Fe-pnictides near magnetism in compounds that allow comparison of antimonides with arsenides and phosphides. However, also different from the Fe-based superconductors, which are based on nominally divalent Fe tetrahedrally coordinated by pnictogens, filled skutterudites have a more covalent Fe with octahedral coordination. Further, the filled skutterudites are cubic, while the Fe-based superconductors have layered structures. This is an important difference from the point of view of quantum fluctuations, since dimensionality it thought to play an important role in the behavior of quantum critical systems[34]. Nonetheless, these filled skutterudites do show a composition dependent cross-over between a paramagnetic metal to a magnetically ordered state with signatures of itinerant character and perhaps sizable effects due to quantum fluctuations in the cross-over region[35–37]. These include transport signatures, highly enhanced specific heats, strong orderings induced by rare-earth moments, and magnetotransport signatures.

The purpose of this paper is to elucidate the chemical range of the magnetism in these compounds and specifically the extension from the well studied antimonides to the arsenides and phosphides. We also report results for the binary skutterudite, $FeSb_3$ motivated by recent experimental results[38], in relation to reported DFT calculations[39].

## Results

The band structures (Fig. 1) show rough similarity between the La-filled and Na-filled compounds within a family with the exception of the placement of the Fermi energy. The filled skutterudite band structures show a light pnictogen derived band. This light band comes from the band that crosses the gap and forms the top valence band in the unfilled compounds[40,41]. The lower bands are heavier and have substantial Fe 3$d$ contributions. It is these heavy Fe 3$d$ derived bands that lead to magnetism, as was discussed previously[28,30,37]. The different families (P, As and Sb) differ in the shape and placement of the light band with respect to the heavier bands and in the details of the hybridization of the Fe and pnictogen states. This is also seen in the density of states (Fig. 2). However, the details of the band structures of the Na and La compounds are not simply related by rigid band behavior. This is because the position of the light band is influenced by the charge on the filling atom[42], and in particular the light band is lower in energy, closer to the heavy bands, in the La filled compounds, corresponding to the higher charge on trivalent La. This then affects the shape of the density of states in a way that can work against the effect of the change in band filling as regards magnetism (see below).

The relative placement of the light and heavy bands plays a central role in the thermoelectric performance of $p$-type filled skutterudites[41], which has been one of the main interests in the study of filled skutterudites. In this regard, the light band extends higher above the heavy bands in the phosphide skutterudites, which makes them less favorable as thermoelectrics[43]. Perhaps for this reason the phosphide and arsenide filled skutterudites have been less studied as thermoelectrics than the antimonides. This also has consequences for magnetism.

Within Stoner theory[44–46], an instability towards ferromagnetism occurs in the mean field when the Stoner criterion $N(E_F) \geq I^{-1}$ is met. Here $N(E_F)$ is the electronic density of states at the Fermi energy, $E_F$, and $I$ is the Stoner parameter, which depends on the chemical details, but for an Fe pnictide compound is governed by the Fe contribution because the pnictogen $p$ orbitals are very spatially extended, and therefore have a small Hund's coupling. Thus the criterion is roughly $N(E_F)(d) \geq I_{Fe}^{-1}$, where $N(E_F)(d)$ is the Fe 3$d$ contribution to $N(E_F)$ on a per spin and per atom basis, and $I_{Fe}$, which measures the Hunds coupling on Fe, is ~1 eV[47,48].

Therefore a magnetic instability at mean field can be expected when the Fe $d$ projected density of states at $E_F$ significantly exceeds 2 eV$^{-1}$ (i.e. 1 eV$^{-1}$ per spin). This threshold is indicated by the dotted line in the density of states projection (lower panel of Fig. 2). The size of the magnetic instability (i.e. the moments) is controlled by the width of the energy region in the density of states, $N(E)$ over which $N(E)$ is large[45,46]. Turning to the shape of the density of states, there are two competing effects when the charge on the filling atom is reduced going from La to Na. First of all, $E_F$ is lowered which raises $N(E_F)$ due to the heavy bands. The second is that the light band rises relative to the heavy bands, which works against this.

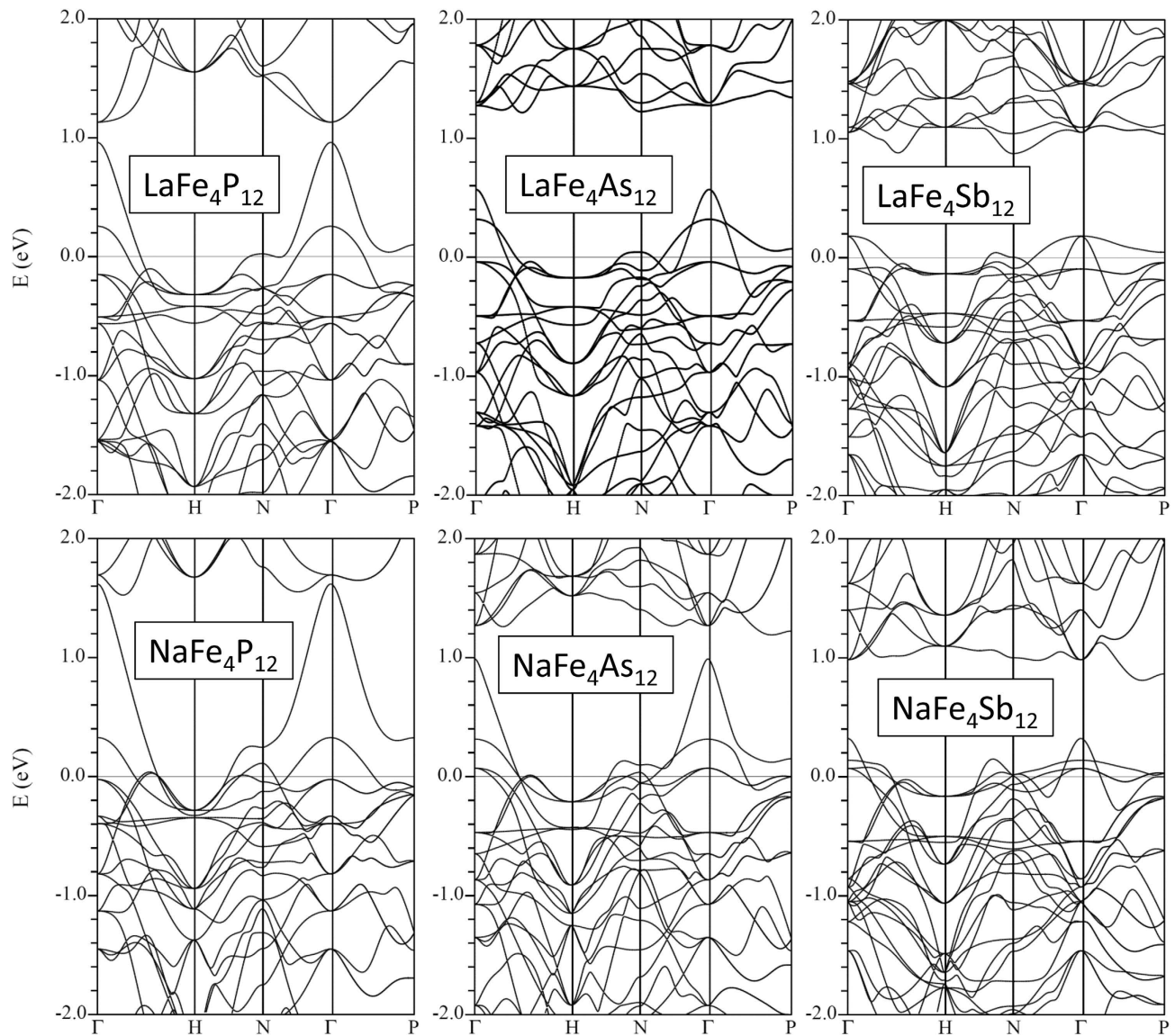

**Figure 1.** Band structures of the various compounds. The Fermi energies are at 0 eV. Note that in all cases the Na compounds have Fermi levels cutting heavy transition element derived bands.

The calculated values of $N(E_F)$ are 26.3 eV$^{-1}$, 43.8 eV$^{-1}$ and 29.9 eV$^{-1}$ on a per formula unit basis for the phosphide, arsenide and antimonide, respectively. The corresponding contributions from Fe $d$ states as measured by the $d$ projection onto an Fe LAPW sphere are 4.1 eV$^{-1}$, 6.7 eV$^{-1}$ and 3.5 eV$^{-1}$ per atom. These values are clearly above the Stoner criterion, and therefore all these Na-filled skutterudites are expected to have ferromagnetic instabilities at the mean field-like DFT level. This is reflected in our self consistent calculations, which show ferromagnetic ground states at the DFT level for all compounds.

We emphasize that in general, while the presence of a mean field ferromagnetic instability is governed by the value of $N(E_F)$ in Stoner theory, once there is an instability, its size is not determined by the value of $N(E_F)$ but on the shape of the density of states in the energy range around $E_F$, as described by extended Stoner theory[45,46]. In particular, a larger instability often results from a broader density of states peak, even though the peak $N(E_F)$ is lower.

Fig. 3 shows the results of fixed spin moment calculations. Fixed spin moment calculations for NaFe$_4$Sb$_{12}$ were reported previously by Leithe-Jaspers and co-workers[28]. As shown, the Na-filled skutterudites are all ferromagnetic at mean field. However, the Sb compound is much more ferromagnetic in terms of both the magnetic energy and the size of the predicted moment (3.0 $\mu_B$ per formula unit for NaFe$_4$Sb$_{12}$, close to 3.0 $\mu_B$ for NaFe$_4$As$_{12}$, and 1.6 $\mu_B$ for NaFe$_4$P$_{12}$). The As compound is intermediate, but much more similar to the antimonide than to the phosphide. We also did unconstrained calculations including spin-orbit for $NaFe_4Sb_{12}$ and $NaFe_4P_{12}$, and obtained very similar results, specifically spin moments of $2.94\,\mu_B$ and $1.62\,\mu_B$, respectively.

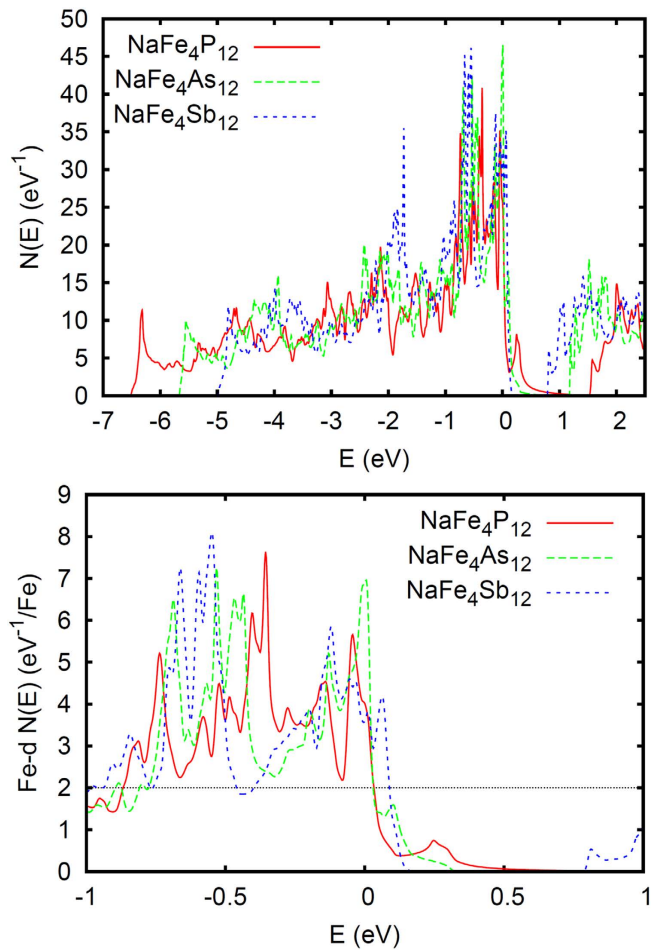

**Figure 2.** Total (top) and Fe-d projections (bottom) of the density of states of NaFe4P12, NaFe4As12 and NaFe4Sb12. The Fe-d projections are onto LAPW sphere radii of 2.1 Bohr for Fe. The Fermi energy is at 0 eV. The dotted line in the bottom panel marks $2\,eV^{-1}$ (see text).

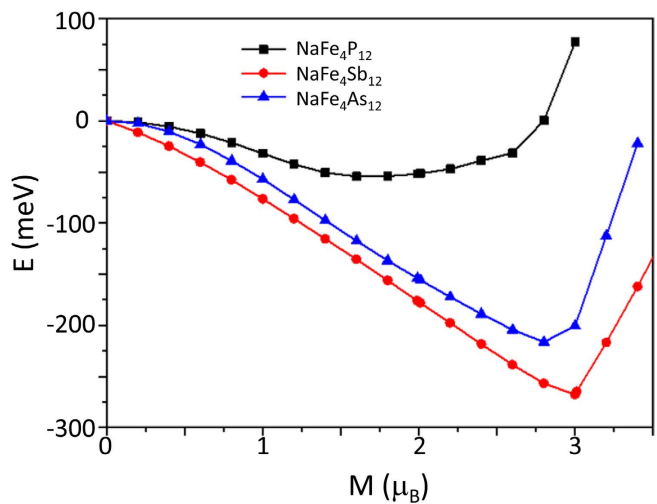

**Figure 3.** Fixed spin moment total energy as a function of constrained spin magnetization as obtained using the PBE GGA for NaFe4P12, NaFe4As12 and NaFe4Sb12 on a per formula unit basis. The energy zero is at the energy of the non-spin-polarized case.

Although hardly unusual, it is worth noting that the magnetic energies and magnetizations do not correlate with the size of $N(E_F)$. The reason for the differences is seen in the band structures and corresponding density of states, $N(E)$. As mentioned, the light band is higher in the phosphides than in the arsenide and antimonide filled skutterudites. This leads to a lower $N(E)$ in the phosphides as the energy is raised above $E_F$ along with a wider region of low $N(E)$ near the band edge (3 electrons per formula unit above $E_F$ for the Na-filled compounds). Based on this analysis, $NaFe_4P_{12}$ is expected to be ferromagnetic, but with weaker ferromagnetism than $NaFe_4As_{12}$ and $NaFe_4Sb_{12}$, which is what is found in the full calculations. We also did local spin density approximation (LSDA) calculations for $NaFe_4P_{12}$. We find again a ferromagnetic ground state (spin moment of $1.3\,\mu_B$ per formula unit), supporting the conclusion from the PBE GGA calculations that this compound has a ferromagnetic ground state at the mean field like DFT level.

Mochel and co-workers[38], have reported experimental investigation of the transport and susceptibility of skutterudite structure $FeSb_3$ films. The authors extracted a band gap of 16.3 meV from a fit of the resistivity. The measured resistivity was found to decrease by ~20% with temperature from ~30 K to 300 K. The temperature dependent resistivity data does not, however, properly fit the activated form that would be expected for a semiconductor with a defined gap over any temperature range and furthermore has a maximum at ~30 K and decreases as temperature is lowered below this. The susceptibility does not show any magnetic transition as a function of temperature, but instead nearly diverges as the temperature is lowered, showing Curie-Weiss behavior with a Weiss temperature close to 0 K. These behaviors could be consistent with a material very close to a ferromagnetic instability.

Rasander and co-workers[39], reported detailed lattice dynamics and electronic structure calculations for this phase, finding that the compound has a ferromagnetic ground state with a high estimated Curie temperature $T_c = 175$ K, and that it is only dynamically stable when ferromagnetism is included. The necessity of including magnetism in order to accurately describe the crystal structure even in the non-magnetic state is a characteristic also found in the Fe-based superconductors[14].

We repeated these calculations to confirm this potentially important result in relation to quantum criticality in skutterudites. We did calculations with both the PBE GGA and with the LSDA, which sometimes gives weaker magnetic tendencies than the GGA. We find ferromagnetic instabilities in both cases, which supports the findings of Rasander and co-workers[39]. Specifically, for the optimized PBE structures we obtain a magnetization of $4\,\mu_B$ per unit cell (four formula units of $FeSb_3$) with a magnetic energy of 280 meV. In the LSDA, relaxing the lattice parameter according to the LSDA, we obtain a somewhat smaller magnetization of $3.55\,\mu_B$ per unit cell, with an energy gain of 50 meV. This is a severe test because the LSDA volume is 6.5% smaller than the PBE volume and this compression works against magnetism (the calculated lattice parameters for the ferromagnetic state are 9.175 Å and 8.971 Å for the PBE GGA and the LSDA, respectively, while the corresponding non-spin-polarized lattice parameters are 9.151 Å and 8.960 Å). Thus both the LSDA and PBE functionals predict a ferromagnetic ground state in contrast to the experiment, again suggesting further investigation of this material for further signatures of quantum critical behavior.

## Discussion

To summarize, we find that at the standard density functional level (PBE GGA), $NaFe_4Sb_{12}$ is ferromagnetic in accord with experiment and previously reported calculations. As mentioned, based on experimental data, this compound shows significant effects of quantum fluctuations. In general, such fluctuations weaken the magnetic instability. This effect has been observed in $NaFe_4Sb_{12}$ and $KFe_4Sb_{12}$, where ordered moments of $\sim 1\,\mu_B$ per formula unit are found experimentally instead of $3\,\mu_B$ predicted by density functional calculations[28,29]. This is reminiscent of other itinerant metals near a ferromagnetic quantum critical point.

The resulting renormalization is controlled by the magnetic energy surface (the bare energy as a function of constrained moment) and the size of the quantum fluctuations in the order parameter, which can vary between different materials[12,49,50]. Following the arguments of Aguayo and co-workers for $Ni_3Al$ in relation to $Ni_3Ga$, the lighter bands in $NaFe_4P_{12}$ relative to $NaFe_4Sb_{12}$ might favor weaker quantum fluctuations in the phosphide. In any case, the present results imply that $NaFe_4P_{12}$ is either a ferromagnet or a renormalized paramagnet in which ferromagnetism has been destroyed by spin fluctuations. This is also the case for the hypothetical compound $NaFe_4As_{12}$, which shows stronger magnetism at the PBE GGA level, closer to the behavior of $NaFe_4Sb_{12}$.

It will be of interest to experimentally study the magnetic properties of $NaFe_4P_{12}$ to determine if the compound has ordered ferromagnetism. In either case, transport, specific heat and magnetotransport measurements looking for signatures of quantum fluctuations will be useful. This also applies to the arsenide, $NaFe_4As_{12}$ if it can be made, and to $FeSb_3$, which exists in thin film form.

## Methods

The present calculations for the filled skutterudites were performed within density functional theory using the generalized gradient approximation of Perdew and coworkers (PBE GGA)[51]. We used the experimental lattice parameters of the filled skutterudites, NaFe$_4$P$_{12}$ and NaFe$_4$Sb$_{12}$ and relaxed the atomic position of the pnictogen, which has two free structural parameters. Skutterudite structure NaFe$_4$As$_{12}$ has not been reported in literature. For this hypothetical compound we relaxed both the lattice parameter and the internal As position using the PBE GGA. The calculated lattice parameter for this phase is 8.345 Å. The structural data is given in Table 1.

| Compound | $a$(Å) | $u$ | $v$ | $N(E_F)$(eV$^{-1}$) |
|---|---|---|---|---|
| LaFe$_4$P$_{12}$ | 7.8316 | 0.3531 | 0.1522 | 14.8 |
| LaFe$_4$As$_{12}$ | 8.2152 | 0.3443 | 0.1570 | 14.7 |
| LaFe$_4$Sb$_{12}$ | 9.1395 | 0.3360 | 0.1633 | 19.4 |
| NaFe$_4$P$_{12}$ | 7.782 | 0.3530 | 0.1484 | 26.3 |
| NaFe$_4$As$_{12}$ | 8.345 | 0.3445 | 0.1546 | 43.8 |
| NaFe$_4$Sb$_{12}$ | 9.1767 | 0.3328 | 0.1620 | 29.9 |

**Table 1.** Structures and calculated $N(E_F)$ per formula unit for the various compounds. $u$ and $v$ are the free parameters in the pnictogen position ($u,v$,0).

We used the general potential linearized augmented planewave (LAPW) method[52] as implemented in the WIEN2k code for most of the calculations reported here[53]. We employed well converged basis sets including local orbitals, with the parameter $R_{min}k_{max}=9$ ($R_{min}$ is the minimum LAPW sphere radius and $k_{max}$ is the cutoff for the planewave sector of the basis). We used the precise standard LAPW plus local orbital approach to treat semicore states, as opposed to the so called APW+lo method[54,55]. Dense zone samplings were used, generally $24 \times 24 \times 24$ meshes, and we tested convergence by varying these. We also did calculations with different sphere radii as tests and also repeated some calculations using the projector augmented wave (PAW) method[56,57]. as implemented in the VASP code with standard settings[58,59]. The calculations for skutterudite FeSb$_3$ were done using the VASP code.

## Acknowledgments

Work at ORNL was supported by the Department of Energy, Basic Energy Sciences, Materials Sciences and Engineering Division. YM acknowledges funding support from the Natural Science Foundation of China (Grant No. 11025418).

## Author Contributions

D.J.S., G.X., H.S. and X.F. did first principles calculations. D.J.S., G.X., X.F., W.Z., Y.M. and H.S. contributed to the concept and analysis of results. D.J.S., G.X., X.F., W.Z., Y.M. and H.S. participated in drafting the manuscript.

## Additional Information

**Competing financial interests:** The authors declare no competing financial interests.

**How to cite this article**: Xing, G. *et al.* Magnetism in Na-filled Fe-based skutterudites. *Sci. Rep.* **5,** 10782; doi: 10.1038/srep10782 (2015).